\documentclass[fleqn,gc,epsf,amsfonts]{aipproc}
\usepackage{amssymb,graphicx}
\layoutstyle{8x11single}

\begin{document}

\title{Some impacts of quintessence models on cosmic structure formation}

\author{A. Füzfa$^{1}$, J.-M. Alimi$^{2}$}{
  address={Laboratory Universe and Theories, CNRS UMR 8102,\\
 Observatoire de Paris-Meudon and Universit\'e Paris VII, France\\
           $^{1}$ email: andre.fuzfa@obspm.fr\\
           $^{2}$ email: Jean-michel.alimi@obspm.fr}}


\begin{abstract}
Some physical imprints of quintessence scalar fields on dark matter (DM) clustering are illustrated, and a comparison with 
the concordance model $\Lambda CDM$ is highlighted.
First, we estimate the cosmological parameters for two quintessence models, based
on scalar fields rolling down the Ratra-Peebles or Sugra potential, by a statistical analysis of the Hubble diagram
of type Ia supernovae. 
Then, the effect of these realistic dark energy models on large-scale DM clustering
is established through N-body simulations. Various effects like large-scale distribution of DM,
cluster mass function and halos internal velocities are illustrated. 
It is found that realistic dark energy models lead to quite different DM clustering, due to a combination of the variation of the equation of state and 
differences in the cosmological parameters, even at $z=0$. This conclusion contradicts other works in the recent litterature
and the importance of considering more realistic models in studying the impact of quintessence on structure formation is highlighted.
\end{abstract}

\maketitle

\section{Introduction}
In recent years, there have been increasing evidences in favour of an
unexpected energy component --- often called dark energy
--- which dominates the present universe and affects the recent cosmic
expansion. The first evidence came from the Hubble diagrams of type Ia
supernovae \cite{riess1,perlmutter,riess2,snls} and were thereafter confirmed by
the measurements of
Cosmic Microwave Background (CMB) anisotropies (cf. \cite{bennett}
and the other companion papers), the large-scale distribution of
galaxies \cite{tegmark1,tegmark2} and indirectly by the observed properties of
galaxy clusters \cite{allen} and peculiar velocities \cite{mohayaee}.
This acceleration can be explained by the present domination of an exotic type of matter --- often
called dark energy and being characterised
by a negative pressure --- over all other ordinary types of energy with positive equation of state (eos) $\omega=p/\rho$.
\\
\\
However, several questions have arised on the nature of the underlying physical process behind this acceleration.
The usual and most simple explanation is that the acceleration
is due to a positive cosmological constant, which acts like a fluid with a constant eos $\omega_\Lambda=-1$.
This interpretation leads nevertheless to a few intricate problems. First, the value of the cosmological constant
has to be extremely fine-tuned to account for the observed amount of accelerating energy: $\Lambda\approx 5\times 10^{-84}GeV^2$
for $\Omega_{\Lambda,0}=0.7$ (in natural units: $\hbar=c=1$). We refer the reader to \cite{weinberg} for a review on this fine-tuning problem
and
to \cite{padmanabhan} for an interesting alternate interpretation.
The situation about the cosmological constant has been made even worst by the discovery
of dark energy in recent years because the observational evidences
claims for a value of vacuum energy density $\rho_\Lambda$ of the order of the present critical density of the universe. In
addition to the fine-tuning problem evoked above, the interpretation
of dark energy in terms of a cosmological constant coming from non-vanishing
vacuum energy also faces this coincidence problem to explain this
observed value of $\Lambda$. These problems concerning the
cosmological constant have made the need of different interpretation
of dark energy more and more crucial.
\\
\\
A negative eos $\omega=p/\rho$
can be easily obtained in the framework of scalar fields coupled to gravity. This scenario, often called quintessence, relies
on the non-conformally invariant dynamics of such a neutral scalar field coupled to gravity. An impressive number of such quintessence scenarii 
have been suggested in the past few years, by Ratra \& Peebles \cite{ratra}, Brax \& Martin \cite{brax}, Barreiro, Copeland \& Nunes \cite{barreiro},
Frieman et al. 1995 \cite{frieman} amongst many others.
An accelerating phase can
be easily achieved if the potential energy of the field dominates its kinetic energy, leading to a negative pressure. 
Even better, an appropriate choice of the self-interaction potential for the scalar field yields
to the existence of an attractor which fortunately rules out the fine-tuning problem.
This is known as "\textit{tracking} potentials" (see \cite{steinhardt}).
But the price to be paid for it is the reliance on a hypothetic scalar field ruled by a particular self-interaction potential.
However, although some physical motivations have been presented for such models, based on supergravity, compactified extra-dimensions or
spontaneous symmetry breaking and so on, a definitive, physical, interpretation has still to be provided.\\
\\
\\
Due to those various possibilities, it appears crucial to find a way to trace back one particular
model, for example a given potential, from observations. This question is quite intricate as it has been shown
that present (and even future) data coming from high-redshift supernovae observations allow
too much degeneracy between very different eos (see \cite{dipietro,caresia}).
Data coming from supernovae cannot therefore discriminate alone different quintessence scenarii and even different analytical approximations
of the evolution of the eos $\omega(z)$, at such low redshifts ($z<1.5$). 
Such an analysis has therefore to be completed by including physical informations at higher redshifts in order to enlighten the
dark energy's obscure details. 
In \cite{brax2}, the authors have performed an exhaustive study of CMB anisotropies within the framework
of two well-known tracking quintessential models with the Ratra-Peebles \cite{ratra} (RP) and Sugra \cite{brax} potentials.
They found that such models are fully compatible with BOOMERanG and MAXIMA-1 data. Indeed, the dark energy was almost
negligible at the recombination epoch ($z\approx 10^3$) so that most of its effects are concentrated in post-recombination anisotropies. This yields to
slight effects in the CMB anisotropies, which are finally almost insensitive to a dependance on the redshift 
of the eos $\omega(z)$ at recent epochs.\\
\\
Large-scale structures formation has been considered with a growing interest in recent years for the dark energy debate
because of intermediate redshifts ($0<z<1000$) this process
covers. As early as in 1999, Ma and his
collaborators \cite{ma} performed N-body simulations to determine the mass power spectrum with an effective dark energy
with a constant eos throughout the cosmic expansion. Other works by
Bode \cite{bode}, Lokas \cite{lokas}, Munshi et al. \cite{munshi} concentrated on
cluster mass functions but all proceeded with the same,
rather restrictive, assumption of a constant eos.
However, analytical works
on the linear and second order growth rate of the fluctuations as well as the shape of non-linear power spectrum
using analytical approximations were performed by Benabed \& Bernardeau in \cite{benabed} with RP and Sugra quintessence models. 
More recently, Dolag and his collaborators \cite{dolag} studied
numerically the concentration parameters of dark matter halos and their related evolution in the framework of several dark energy models
with both constant and
evolving eos (RP and Sugra models). 
Klypin and his collaborators \cite{klypin} also performed N-body simulations on both
constant and evolving eos (RP and Sugra).
The last found slight differences in
the mass power spectrum and the cluster mass function at $z=0$ that render them almost undistinguishable from
the predictions of a standard $\Lambda CDM$ model. They also discovered more significative differences 
at higher redshifts in those quantities
as well as different halo profiles at all redshifts. However, they based their analysis solely on models with same 
cosmological parameters and in poor agreement with supernovae data as we will show thereafter.
In \cite{solevi}, the authors used the same cosmological models as in \cite{klypin} to study galaxy distribution as
a test of varying eos. They found again large differences at large redshift between the dark energy models
which are reduced substantially by geometrical factors. They concluded that the galaxy redshift distribution should be used 
as a test of varying eos instead of the cluster redshift distribution which leads to too slight differences.
However, we will show here that considering realistic quintessence models --- with different cosmological parameters 
selected by a Hubble diagram analysis --- enlarges the discrepancies 
due to the different cosmic evolutions, even at $z=0$. 
\section{Accelerated cosmological models}
\begin{figure}
\begin{tabular}{ccc}
\includegraphics[scale=0.3]{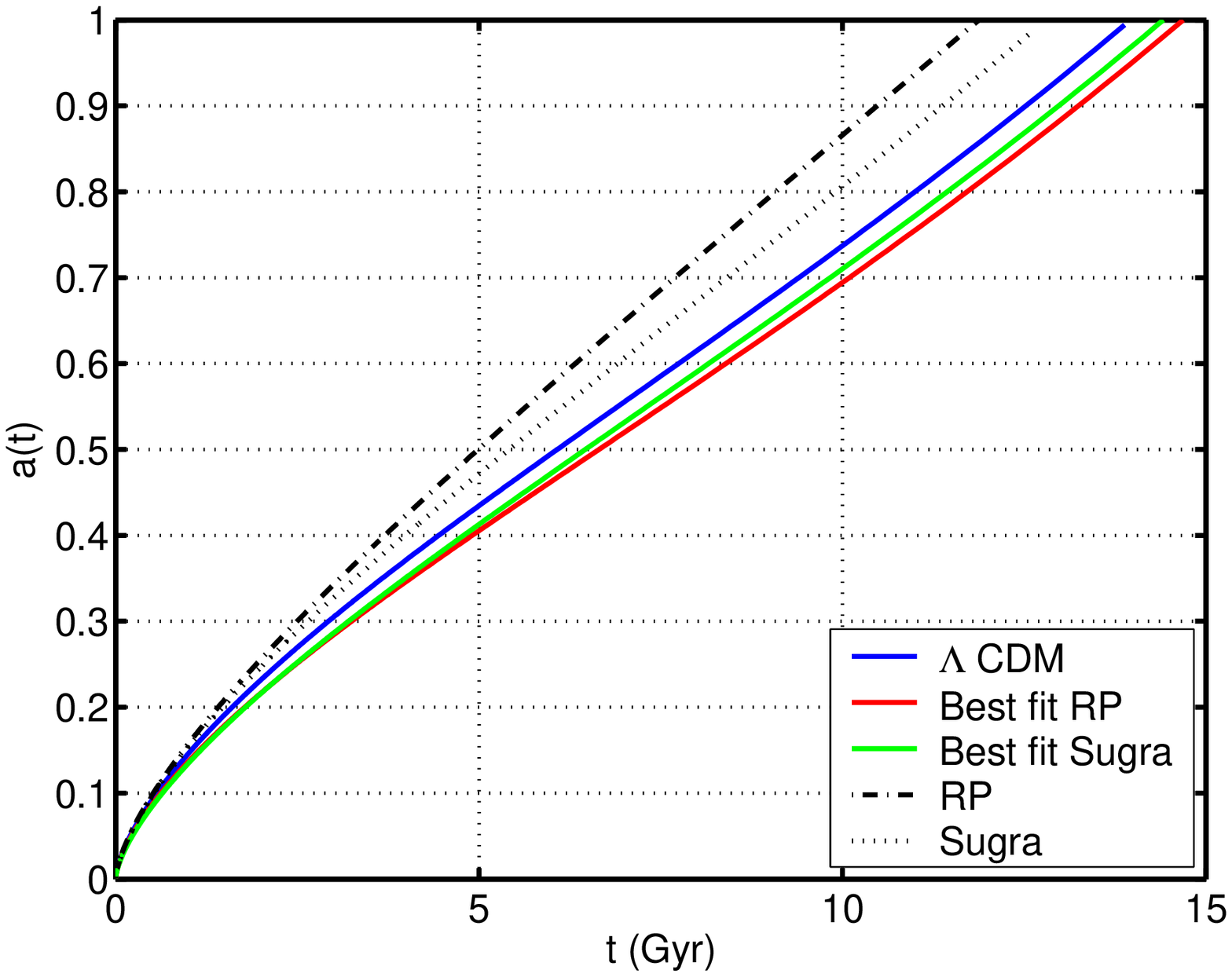} &
\includegraphics[scale=0.3]{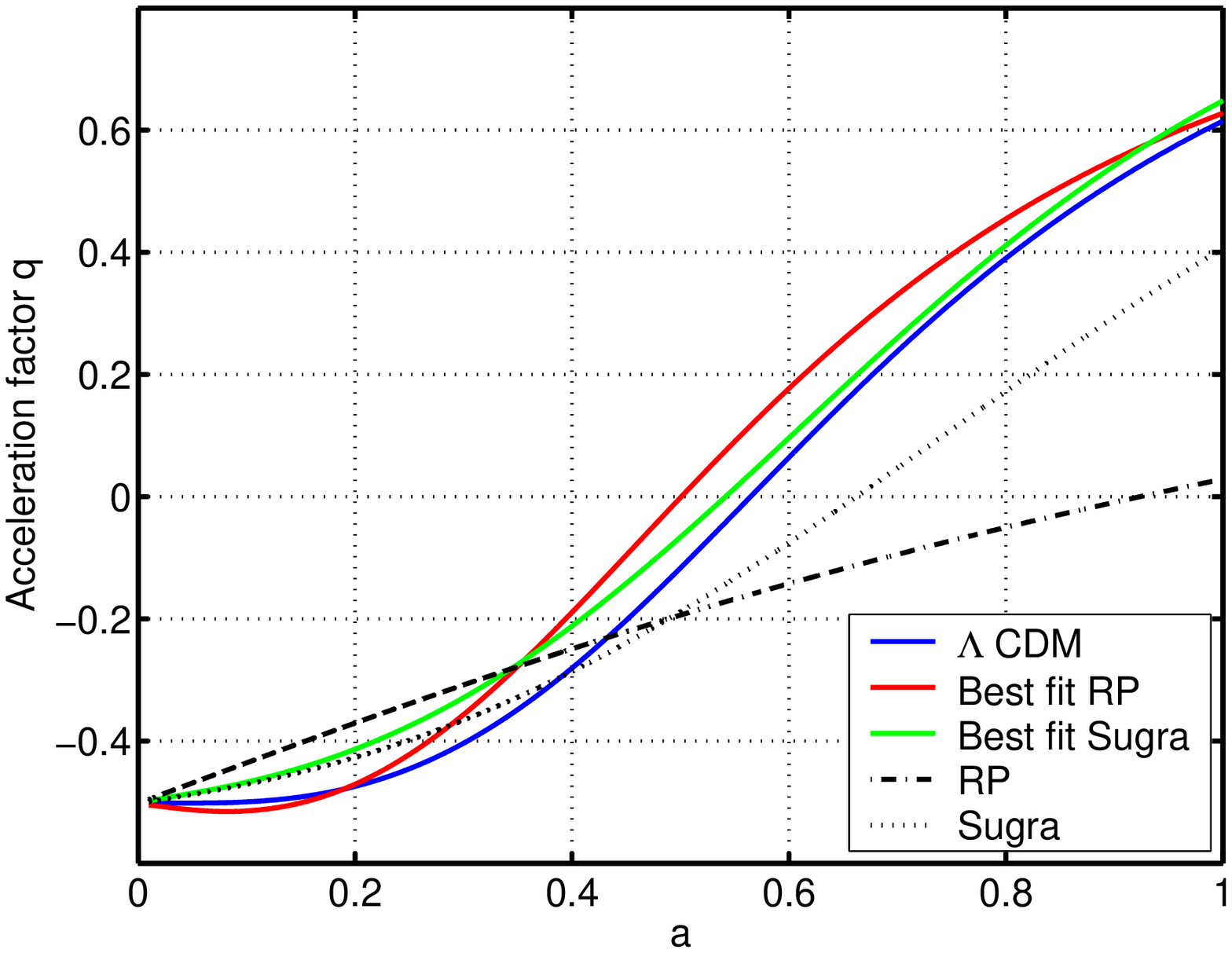} &
\includegraphics[scale=0.3]{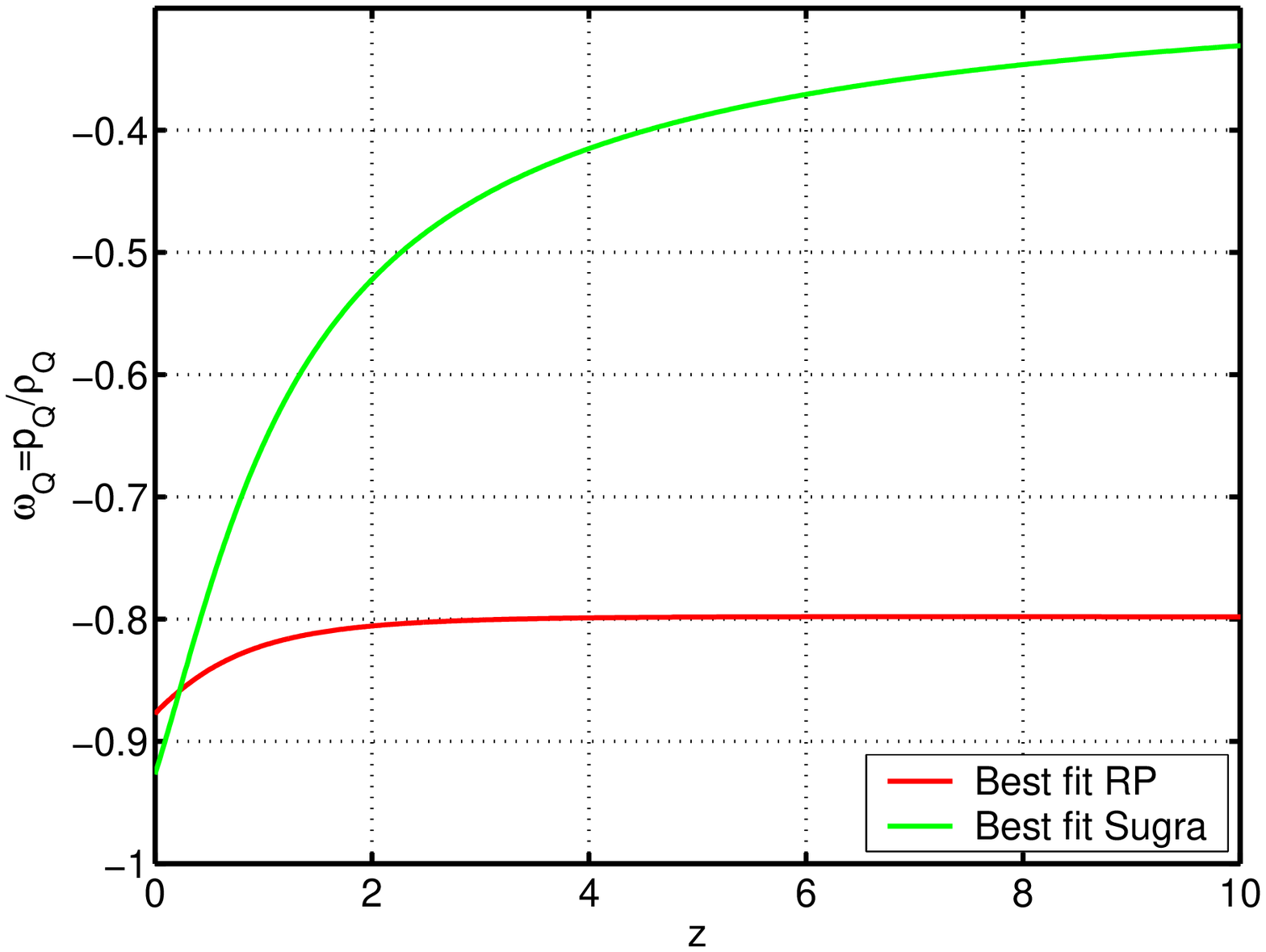}
\end{tabular}
\caption{Cosmological evolutions of the scale factor (left), acceleration factor (center) $q=\ddot{a}a/\dot{a}^2$
and eos $\omega_Q=p_Q/\rho_Q$
for the models in table \ref{chis}. The toy models used in \cite{klypin,solevi} are given in black curves}
\label{models}
\end{figure}

\begin{figure}
\begin{tabular}{cc}
\includegraphics[scale=0.3]{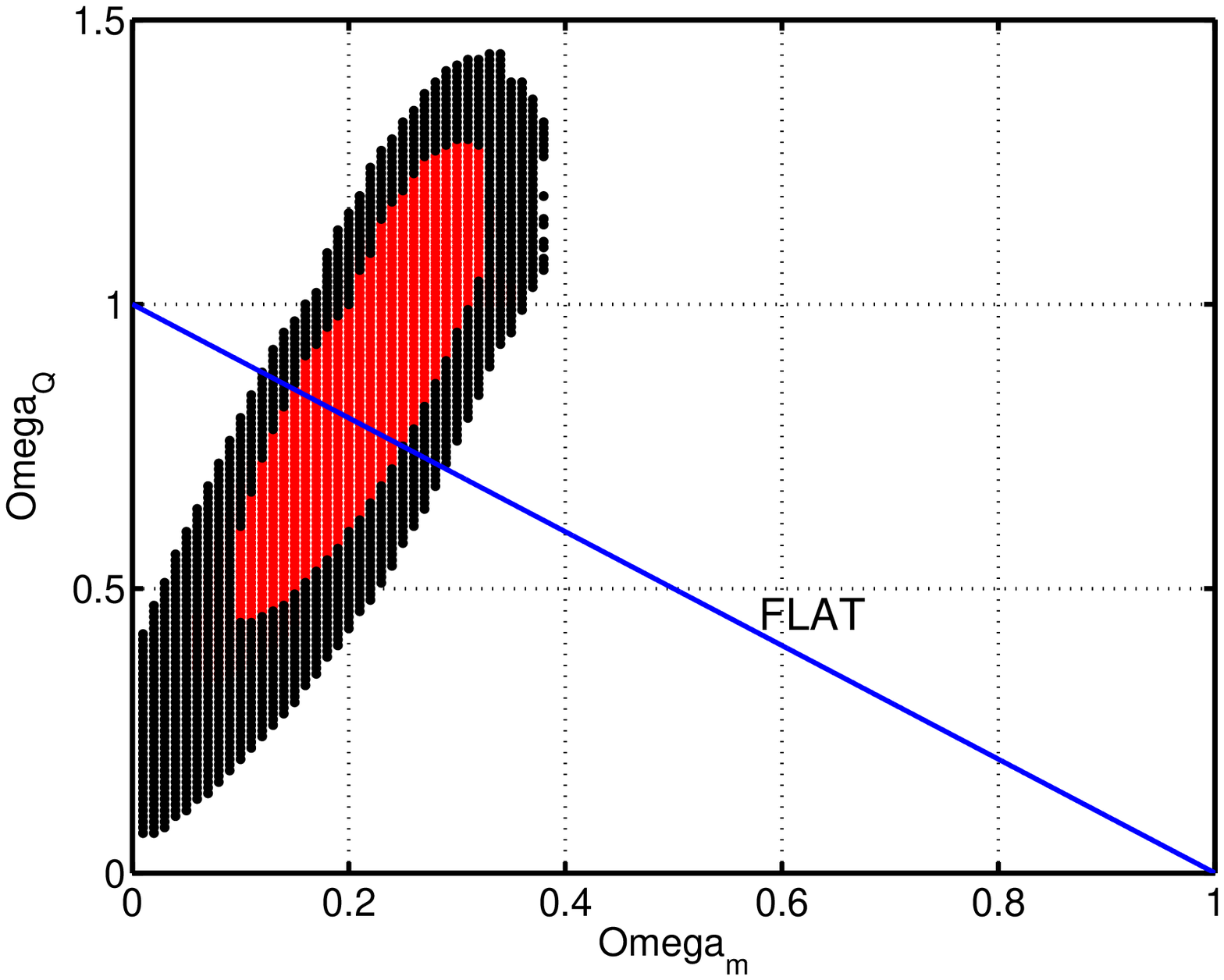} &
\includegraphics[scale=0.3]{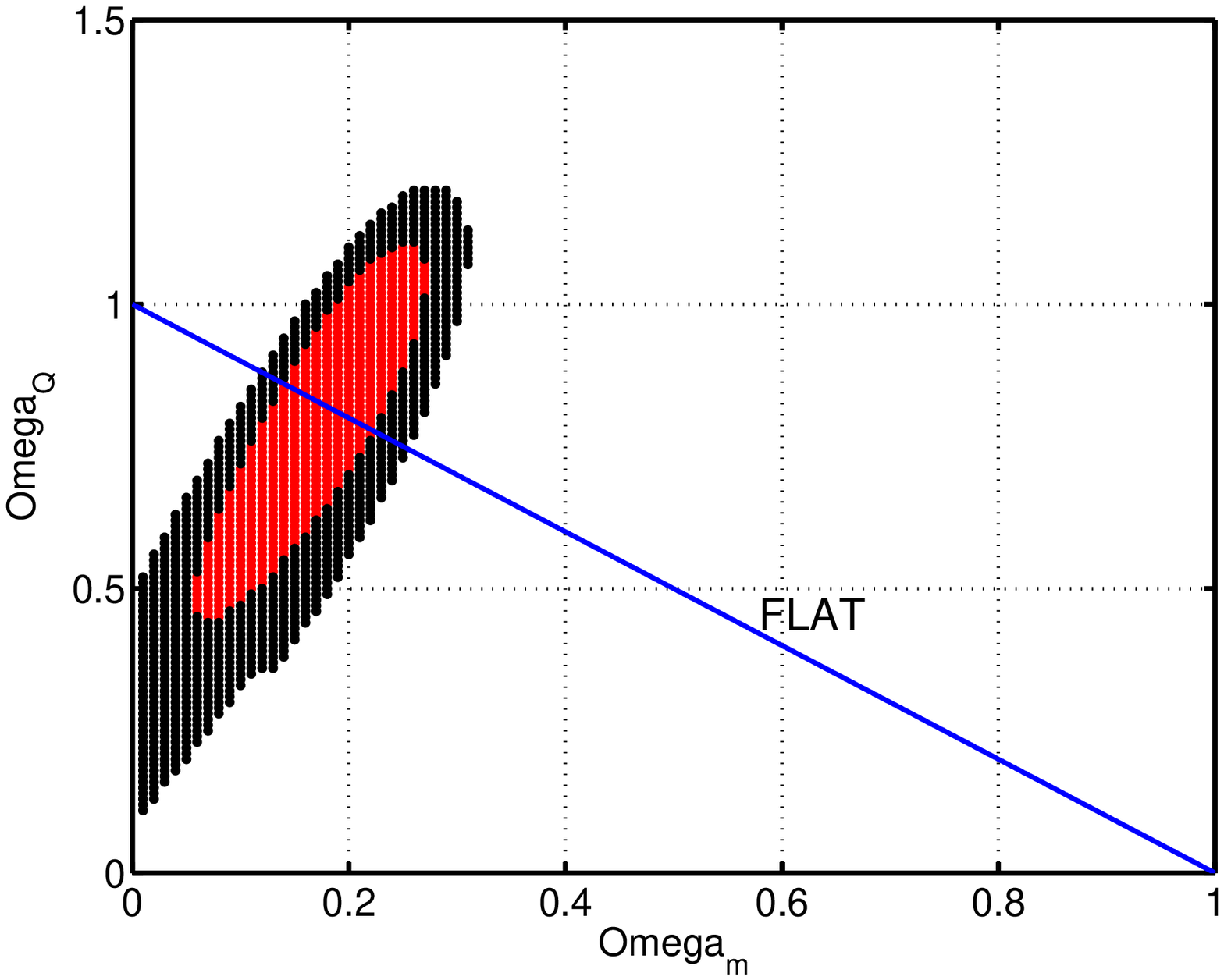}
\end{tabular}
\caption{$68$ (red) and $95\%$ (black) confidence contours for the RP (left) and Sugra (right) models.}
\label{confregions}
\end{figure}

According to the cosmological principle, the large-scale Universe is roughly homogeneous and isotropic and can thus be well
described by the
Friedmann-Lema\^{i}tre-Robertson-Walker (FLRW) line element:
\begin{equation}
\label{flrw}
ds^2=dt^2-a^2(t)\left(\frac{dr^2}{1-k\;r^2}+r^2d\Omega^2\right),
\end{equation}
where $a(t)$ is the scale factor, $k$ is the signature of the curvature and $d\Omega$ is the solid angle element.
We will make use throughout
this paper of the Planck units system in which $\hbar=c=1$, $m_{Pl}=1/\sqrt{G}=1.2211\times 10^{19} GeV$.
The Universe is assumed to be filled with different matter fluids --- pressureless matter (composed of ordinary 
baryonic matter and cold dark matter), relativistic matter (photons and neutrinos) and the mysterious fifth component, called quintessence,
that is hoped to account for the observed cosmic acceleration. The expansion rate of the Universe, characterized
by the \textit{Hubble parameter} $H=\dot{a}/a$, is dictated by the Friedmann equation:
\begin{equation}
\label{friedmann}
H^2=H_0^2\left(\Omega_{m,0}\frac{a_0^3}{a^3}+\Omega_{r, 0}\frac{a_0^4}{a^4}-\Omega_{K,0}\frac{a_0^2}{a^2}+\Omega_{\Lambda,0}\right)+\frac{8\pi}{3m_{Pl}^2}\rho_Q,
\end{equation}
where the subscript zero stands for nowadays value and where the parameters $\Omega_i=\rho_i/\rho_c$ represent the density of the 
different components of the
matter fluids, expressed in critical units. 
More precisely, the subscripts $m$, $r$, $K$, $\Lambda$ and $Q$ indicate respectively the contributions
of the pressureless matter, radiation, curvature, cosmological constant and quintessence to the cosmological expansion.
In equation (\ref{friedmann}),
$\rho_Q$ stands for the density of the quintessence fluid for which there is, in general, 
no analytical expression in terms of the scale factor $a$. For the sake of completeness, we also remind the reader about the acceleration
equation:
\begin{equation}
2\frac{\ddot{a}}{a}=-H_0^2\left(\Omega_{m,0}\frac{a_0^3}{a^3}+2\Omega_{r, 0}\frac{a_0^4}{a^4}\right)-\frac{8\pi}{3m_{Pl}^2}\rho_Q\left(1+3\omega_Q\right),
\end{equation}
where $\omega_Q$ is the ratio between the energy density $\rho_Q$ and pressure $p_Q$ of the quintessence fluid.
\\
\\
We will assume here that the quintessence fluid is actually constituted by a neutral (real) scalar field $Q(t)$ which couples only to ordinary matter
through its gravitational influence (minimally coupled scalar field). The dynamics of such field is ruled by the \textit{Klein-Gordon} equation written in the metric
(\ref{flrw}):
\begin{equation}
\label{kg}
\ddot{\phi}+3\frac{\dot{a}}{a}\dot{\phi}+\frac{1}{m_{Pl}^2}\frac{dV(\phi)}{d\phi}=0,
\end{equation}
where $\phi$ is the field expressed in units of the Planck mass ($Q(t)=m_{Pl}\phi(t)$), and $V(\phi)$ is the field's self-interaction potential. 
The coupling of the quintessence field to gravity is achieved
through its energy density and pressure:
\begin{eqnarray}
\rho_Q&=&\frac{m_{Pl}^2}{2}\dot{\phi}^2+V(\phi),\nonumber\\
p_Q&=&\omega_Q \rho_Q=\frac{m_{Pl}^2}{2}\dot{\phi}^2-V(\phi)\cdot\nonumber
\end{eqnarray}
\\
At this stage, the coupled dynamics of gravitation and the quintessence field is only determined by the choice of a particular potential and initial
conditions, i.e. a couple $(Q_i,\; \dot{Q}_i)$, in the early universe, for example at the end of inflation ($a\approx 10^{-28}$). 
Convenient choices of self-interactions
are those of so-called \textit{tracking potentials} (see \cite{steinhardt} and references therein) that account
for the observed cosmological parameters today from a huge range of initial conditions, spread over dozens of order of magnitudes, 
in the field phase space.\\
\\
Two well-known examples of such potentials are the \textit{Ratra-Peebles} (RP) inverse power law \cite{ratra},
\begin{equation}
\label{rp}
V_{RP}(\phi)=\frac{\lambda^{4+\alpha}}{\phi^\alpha},
\end{equation}
where $\alpha\ge 0$ and $\lambda$ are free parameters,
and the same potential affected by a radiative correction in supergravity due to Brax \& Martin \cite{brax}:
\begin{equation}
\label{sugra}
V_{S}(\phi)=\frac{\lambda^{4+\alpha}}{\phi^\alpha}e^{4\pi\phi^2},
\end{equation}
which we will refer further to the \textit{Sugra} model. The RP model was originally suggested to mimic
a time-varying cosmological constant, which is actually accomplished when the scalar field is frozen by the
expansion at a high energy scale, typically the Planck mass, where the potential is nearly flat. 
However, as Brax \& Martin noticed in \cite{brax},
this state of the field should correspond to a supergravity regime ($Q(t)\approx m_{Pl}$) 
and the potential has therefore to be corrected accordingly.\\
\\
The particular shape in inverse power law of these potentials is hoped to be justified by energy physics (see for example  \cite{brax} and
references therein). In this scenario of quintessence, the field energy density remains subdominant in most of the cosmic
evolution, especially during the radiation dominated era, until we reach the coincidence epoch, which depends exclusively
on the potential shape, due to the tracking property. \\
\\
Let us now estimate the cosmological parameters $\Omega_{m,0}$ and $\Omega_{Q,0}$ to be used for quintessence models
by analysing Hubble diagrams of a type Ia supernovae data set from the SNLS collaboration \cite{snls}.
The moduli distance versus redshift relation of standard candles is given by
\begin{equation}
\label{muz} \mu(z)=m-M=5\log_{10} d_L(z),
\end{equation}
where $d_L(z)$ is the luminous distance
(in Mpc) given by
\begin{eqnarray}
d_L(z)&=&(1+z)H_0\Omega_{K,0}^{-\frac{1}{2}}\sin\left(\Omega_{K,0}^{\frac{1}{2}}\int_0^z \frac{dz'}{E(z')}\right)\;  ; if\; \Omega_{K,0}>0 \nonumber\\
&=&(1+z)H_0\int_0^z \frac{dz'}{E(z')}\; ; if\; \Omega_{K,0}=0 \label{dl}\\
&=&(1+z)H_0|\Omega_{K,0}|^{-\frac{1}{2}}\sinh\left(|\Omega_{K,0}|^{\frac{1}{2}}\int_0^z \frac{dz'}{E(z')}\right)\; ; if\; \Omega_{K,0}<0 \nonumber\\
\end{eqnarray}
for a closed, flat and open FLRW background respectively \cite{peebles} ($E(z)=H(z)/H_0$ is the dimensionless Hubble parameter). 
In order to avoid considering $H_0$ as an additional parameter in the statistical analysis, we marginalize the
$\chi^2$ estimator with respect to this parameter (see also \cite{dipietro}). We will therefore use the following estimator
\cite{dipietro}
\begin{equation}
\bar{\chi}^2=A-\frac{B^2}{C}+\log\left(\frac{C}{2\pi}\right)
\end{equation}
where $A=\sum_{i=1}^{N}\frac{\left(\mu_{obs}(z_i)-\mu_{th}(z_i)\right)^2}{\sigma_i^2}$,
$B=\sum_{i=1}^{N}\frac{\mu_{obs}(z_i)-\mu_{th}(z_i)}{\sigma_i^2}$ and 
$C=\sum_{i=1}^{N}\frac{1}{\sigma_i^2}$ 
with $N$ the number of data, $\sigma_i$ is the uncertainty related to the measurement $\mu_{obs}(z_i)$ and $\mu_{th}$ is the theoretical prediction
for the distance moduli. We choose to fix the value of the parameter $\alpha$ so that $\Omega_{m,0}$ and $\Omega_{Q,0}$ 
will be the only free parameters\footnote{In the case of the Sugra model,
this is well-motivated by the fact that the
exponential correction
in equation (\ref{sugra}) is dominant in the potential
at late redshifts (when $\phi\approx m_{Pl}$), 
leaving the Hubble diagram analysis insensitive to a variation in the parameter $\alpha$.
This was proved \textit{a posteriori} by Caresia and his collaborators (Caresia, Matarrese \& Moscardini, 2004).}. 
The energy scale $\lambda$ in the quintessence potential is here determined 
so that the desired couple $(\Omega_{m,0},\Omega_{Q,0})$  is retrieved.\\
\\
Table \ref{chis} gives the values of $\bar{\chi}^2$ for the best fit RP, Sugra and $\Lambda CDM$ models
as well as the evaluation of this estimator for the quintessence models used in \cite{klypin,solevi}. These last models 
are characterized by an energy scale of $10^3 GeV$ of the quintessence potential (which corresponds to $\alpha\approx 4$) and
the same cosmological parameters. Their agreement with SNLS data is very poor as they lie well outside the $2-\sigma$ confidence
region.
\begin{table}
%%\begin{center}
\begin{tabular}{ccccccc}\hline
Models & $\Omega_{m,0}$ & $\Omega_{Q(\Lambda),0}$ & $\bar{\chi}^2$ & $z_c$ & $z_a$ & $z_{in}$ \\
\hline
RP $\alpha=0.5$ & $0.23$ & $0.91$ & $116$ & $0.71$ & $0.99$ & $33$ \\
\hline
Sugra $\alpha=6$ &$0.2$ & $0.84$ & $116$ &$0.79$ & $0.85$ & $41$  \\
\hline
$\Lambda$ (SNLS) &$0.27$ & $0.75$ & $116$ &$0.41$ & $0.78$ & $37$ \\
\hline
\hline
RP* & $0.3$ & $0.7$ & $168$ &$0.82$ & $0.08$ & - \\
\hline
Sugra* & $0.3$ & $0.7$ & $127$ &$0.43$ & $0.51$ & - \\
\end{tabular}
\caption{\textsl{Cosmological parameters and values of $\chi^2$ on the considered data set for the different models.
The coincidence and acceleration redshifts $z_c$ and $z_a$ as
well as the starting redshift $z_{in}$ of the N-body simulations are also given}}\label{chis}
%\end{center}
\end{table}
Figure \ref{models} represents the cosmological evolution of the models in Table \ref{chis}. 
The age of the Universe lies between $13.9$ ($\Lambda CDM$ model) and $15.9$ (RP)
billion years. Also represented are the models previously used in 
\cite{klypin,solevi} to study the impact of quintessence on structure formation (see also Table \ref{chis}), which are
quite different of the realistic dark energy models used here.
Figure \ref{models} also illustrates the cosmological evolution of the eos $\omega_Q=p_Q/\rho_Q$ for the best fit RP and Sugra
models.
This shows that the same adequacy to data of a Hubble diagram
at low redshifts $z<1-2$ is obtained with very different the eos, leaving
this quantity very difficult to determine from supernovae data alone as already shown in \cite{dipietro}. There is a degeneracy
in the analysis of Hubble diagram as the same cosmic acceleration can be obtained either with a small quantity dark energy ($\Omega_Q$ small) 
providing strong acceleration ($\omega_Q\ll 0$) (for example, this is the case of the cosmological constant here) 
or a larger amount of dark energy exerting weaker acceleration (this is the case of quintessence models here for which $\omega_Q>-1$).
Table \ref{chis} gives the coincidence and acceleration redshifts $z_c$ ($\rho_Q(z_c)=\rho_m(z_c)$) and $z_a$ respectively for the best fit models
of Table \ref{chis}, illustrating
that quintessence models dominate and accelerate earlier than a cosmological constant. As their accelerating power is smaller 
($\omega_Q>-1$) than a cosmological constant, they need more time to provide the same cosmic acceleration observed in supernovae data.
Therefore, the statistical analysis of Hubble diagram in this section leads 
to a prediction of closed
cosmologies ($\Omega_{tot}=\Omega_m+\Omega_Q>1$) for quintessence (see table \ref{chis}). 
This conclusion makes the cross-analysis with CMB data very interesting for quintessence as this could rule out models with
too large $\Omega_{tot}$, but we will leave this point for further studies.
Figure \ref{confregions} represents the $68$ and $95\%$ confidence level contours for the RP and Sugra quintessence models
(a similar figure for the concordance model $\Lambda CDM$ can be found in \cite{snls}). The statistical analysis
has been performed with only two degrees of freedom, the density parameters $\Omega_m$ and $\Omega_Q$,
as the power-law $\alpha$ in (\ref{rp}) and (\ref{sugra}) has been fixed. In the Sugra model, we took $\alpha=6$ so that 
the energy scale $\lambda$ of the potential is of the order of $10^6 GeV$, a scale related to high-energy physics. In this model, 
the exponential correction in (\ref{sugra}) flattens the potential
which allows a good adequacy to data. With such an energy scale $\lambda$ of $10^6 GeV$ ($\alpha\approx 6$),
the RP potential would not be flat enough to provide the required cosmic acceleration and we reduced $\alpha$ to
$0.5$ for the RP model to make the model compatible with data ($\lambda\approx 1eV$). 
We find $\Omega_m<0.38$ ($\Omega_m<0.31$) and $\Omega_Q<1.44$ ($\Omega_Q<1.21$) at the $95\%$ confidence level
for the RP (Sugra) model.
\\
\\
To conclude this section, we state that the statistical analysis of the Hubble diagram of type Ia supernovae
favours closed quintessential universes. In general, more dark energy is needed to provide the same cosmic acceleration
with quintessence scalar field 
than with a cosmological constant. Furthermore, assuming the same cosmological parameters for quintessence and concordance model
will lead to a bad approximation of the recent cosmic expansion (see table \ref{chis} and
Figure \ref{models}). That is why a preliminary analysis is unavoidable
to determine the best density parameters before performing numerical simulations of structure formation that are expected 
to provide additional constraints on dark energy.
We will now use the first three models of table \ref{chis} to evaluate the discrepancies between 
cosmological constant and quintessence on structure formation in dark energy-driven universes. 
\section{Imprints of quintessence on large-scale dark matter clustering}
\begin{figure}
\begin{tabular}{ccc}
\includegraphics[scale=0.4]{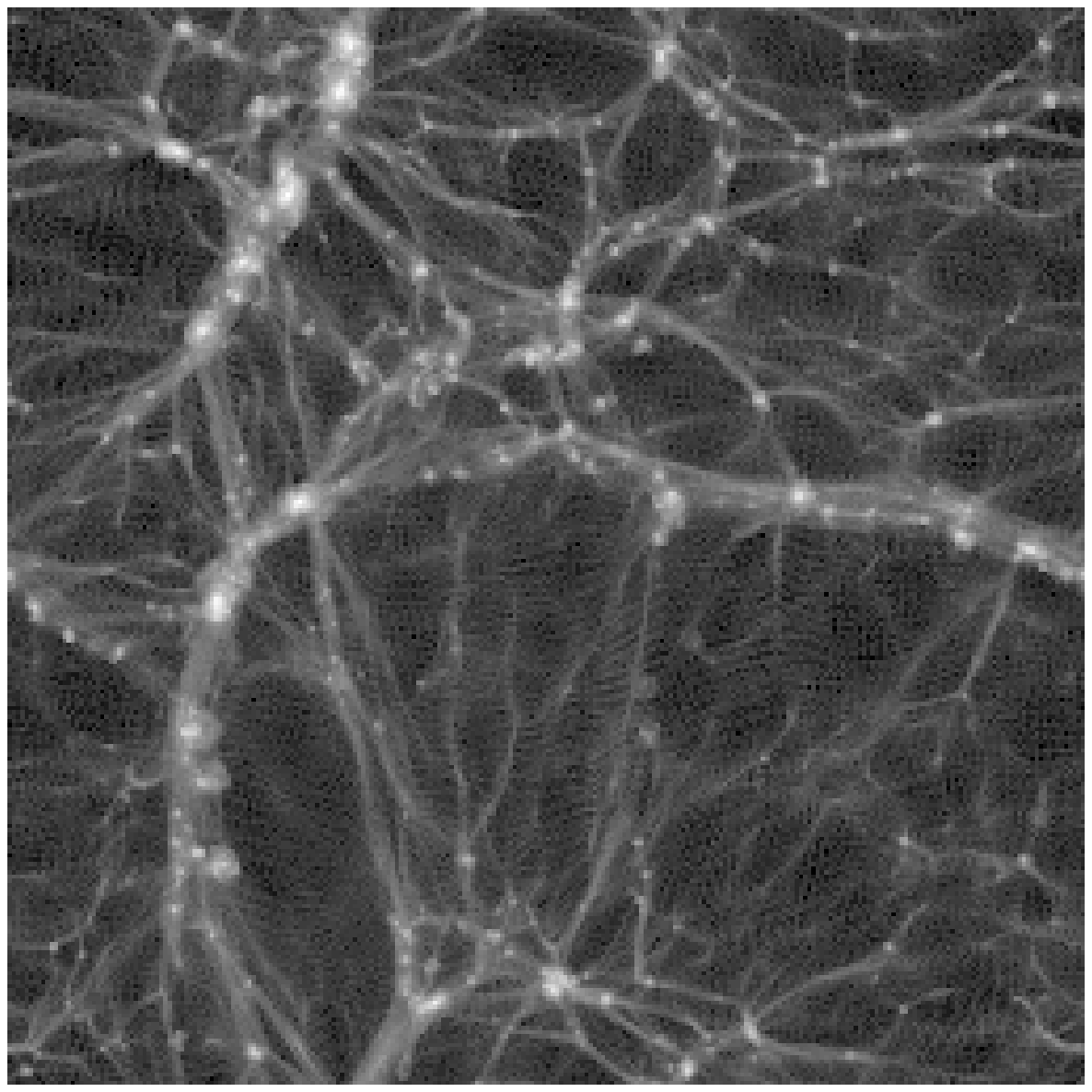} &
\includegraphics[scale=0.4]{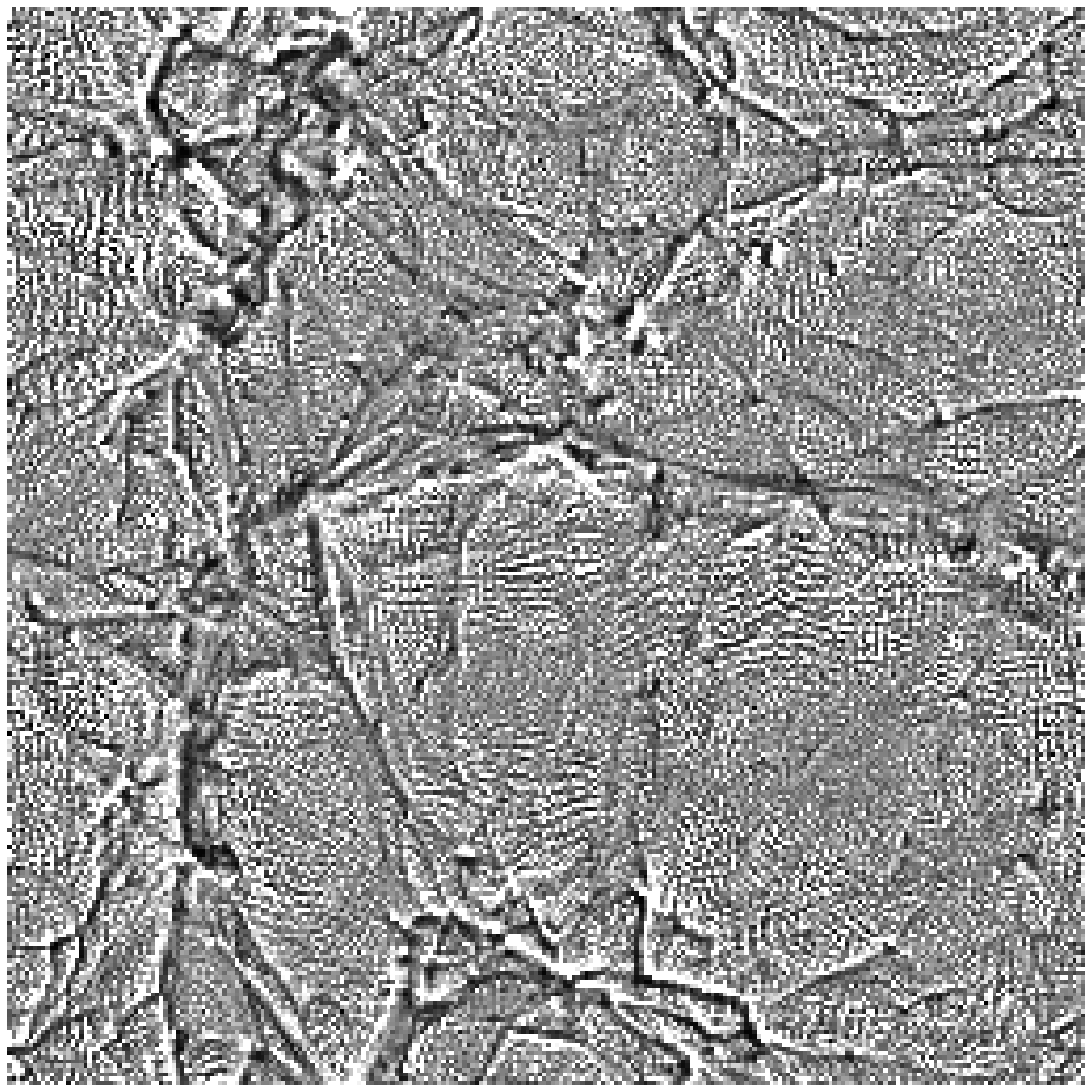} &
\includegraphics[scale=0.38]{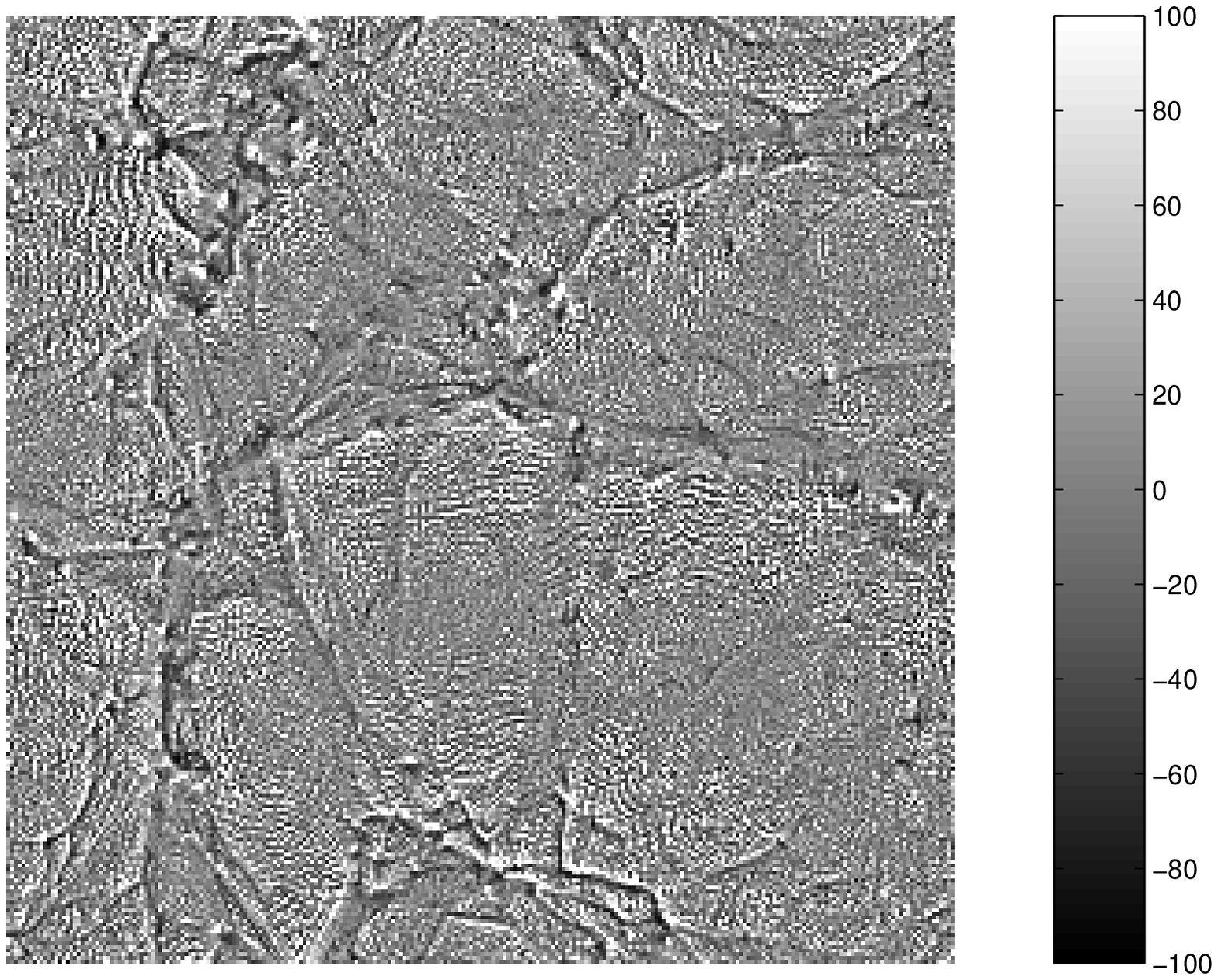}
\end{tabular}
\caption{Dark matter density field in a slice of the $\Lambda CDM$ model (left) and absolute differences for the Sugra (center)
and RP (right) models. The color bar indicate the degree of discrepancy with a $\Lambda CDM$ model in percent.}
\label{slices}
\end{figure}

\begin{figure}
\begin{tabular}{ccc}
\includegraphics[scale=0.3]{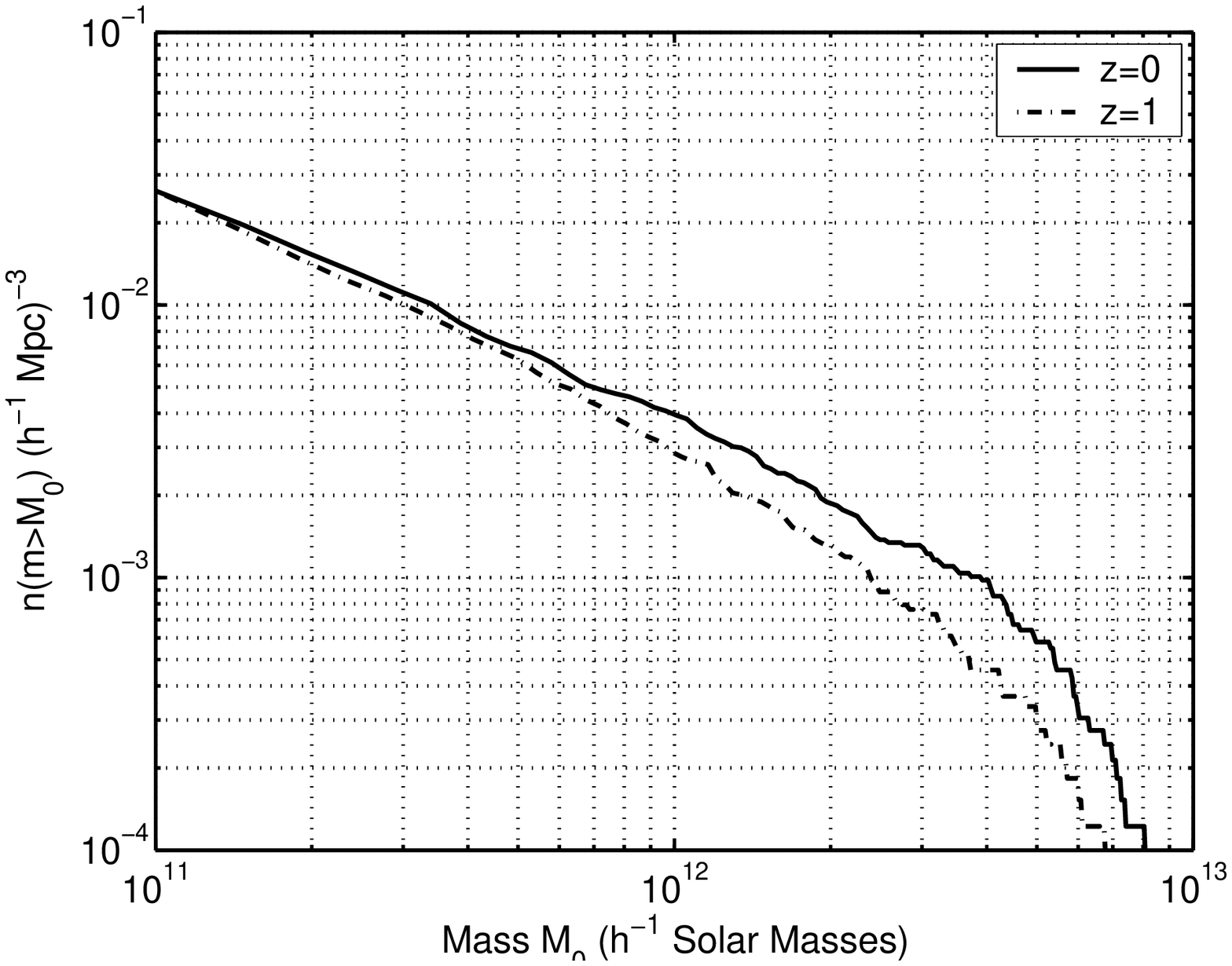} &
\includegraphics[scale=0.3]{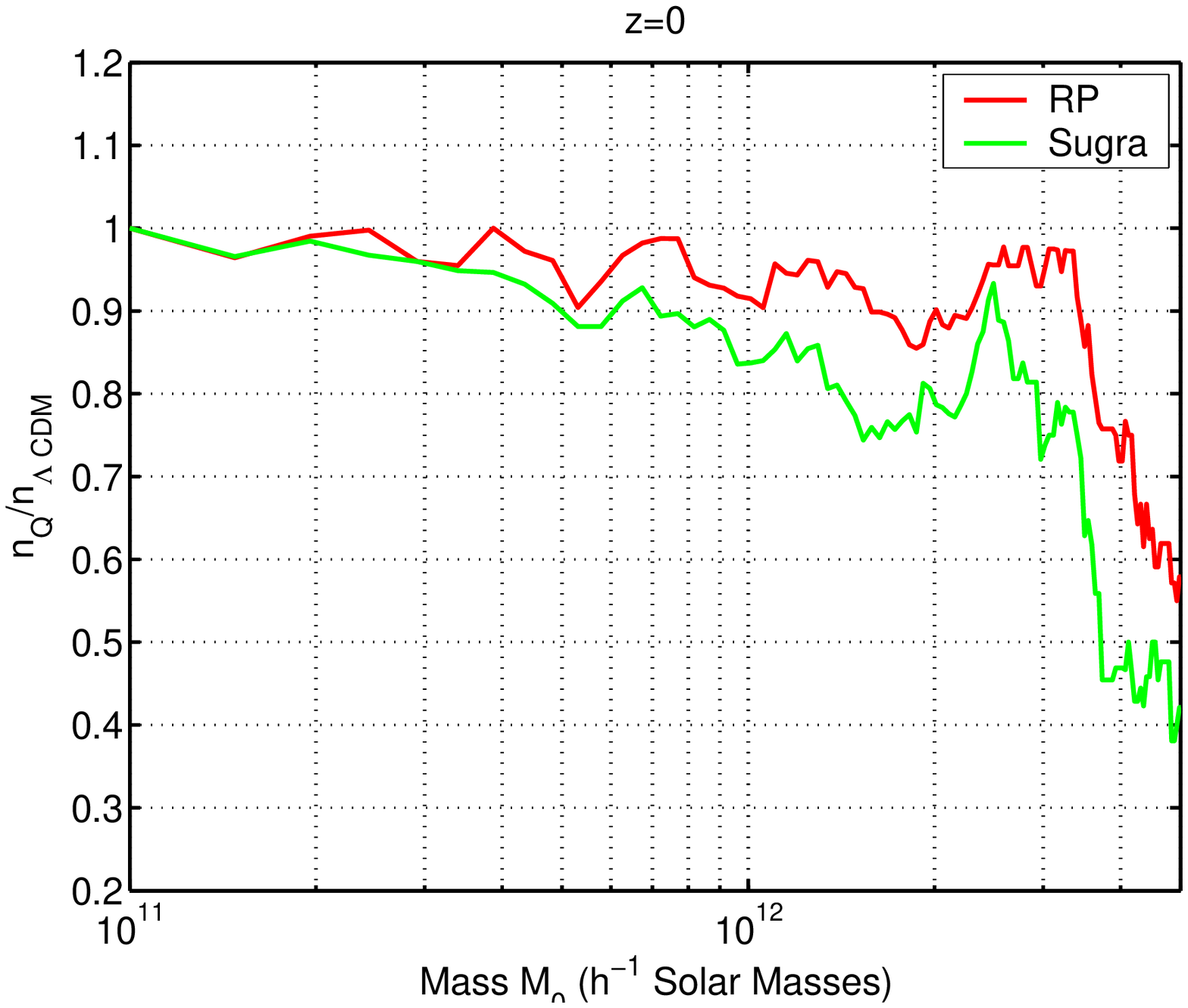} &
\includegraphics[scale=0.3]{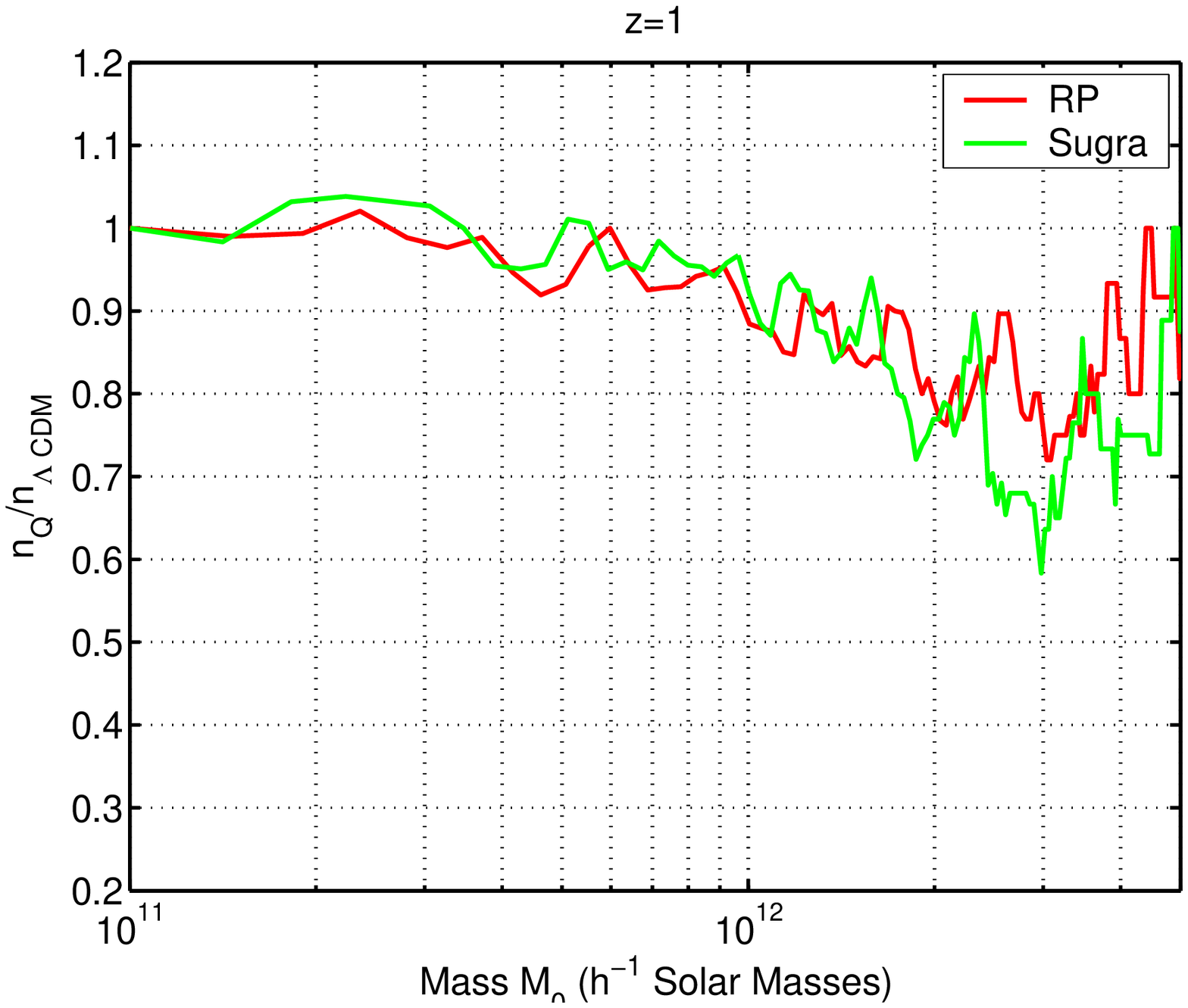}
\end{tabular}
\caption{Mass function of dark matter halos in the $\Lambda CDM$ model at $z=0$ (straight line) and $z=1$ (dash-dotted line) (left).
The mass functions for the quintessence models at $z=0$ (center) and $z=1$ (right) are illustrated  in units of the mass function of the
$\Lambda CDM$ model.}
\label{massfunc}
\end{figure}

\begin{figure}
\begin{tabular}{ccc}
\includegraphics[scale=0.3]{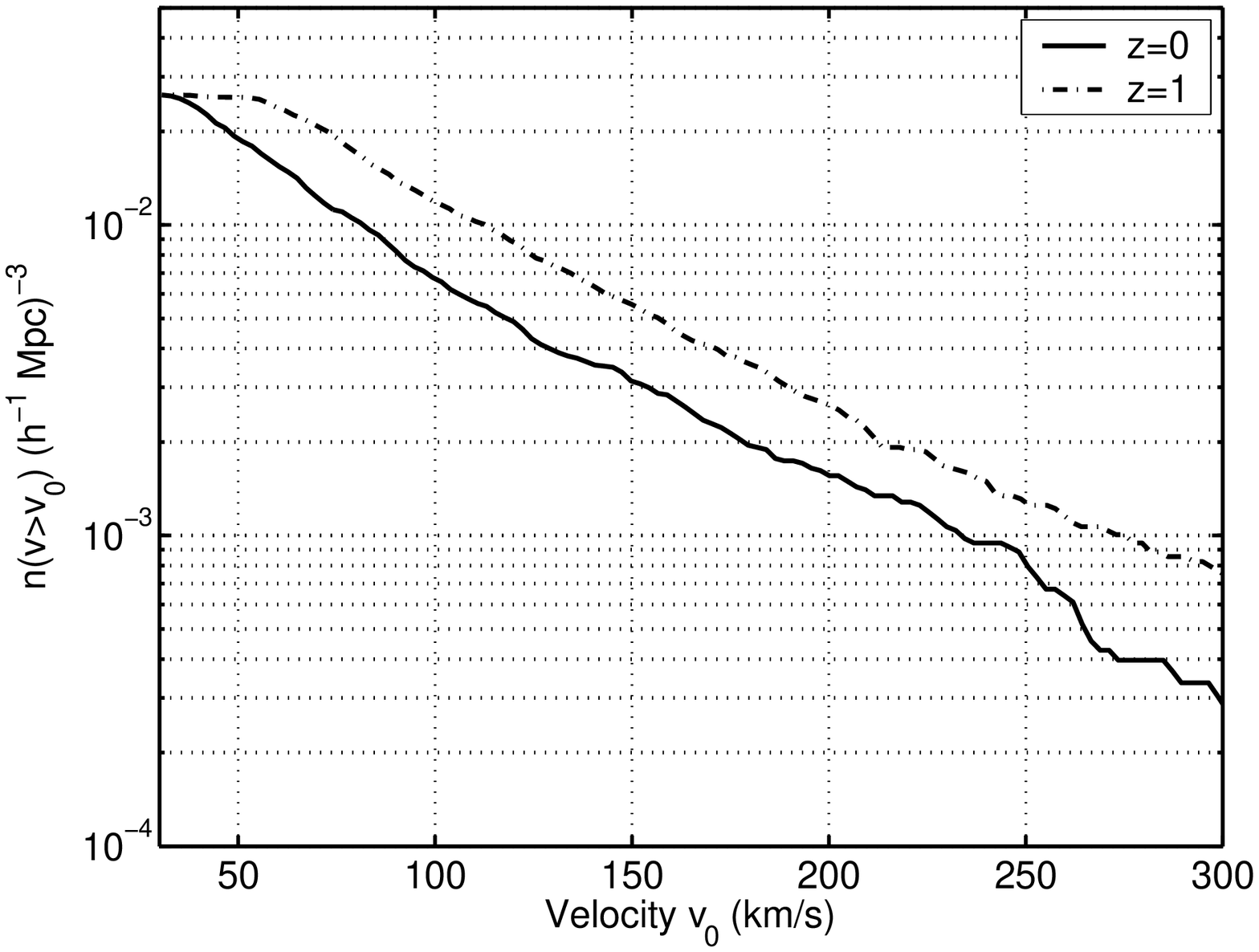} &
\includegraphics[scale=0.3]{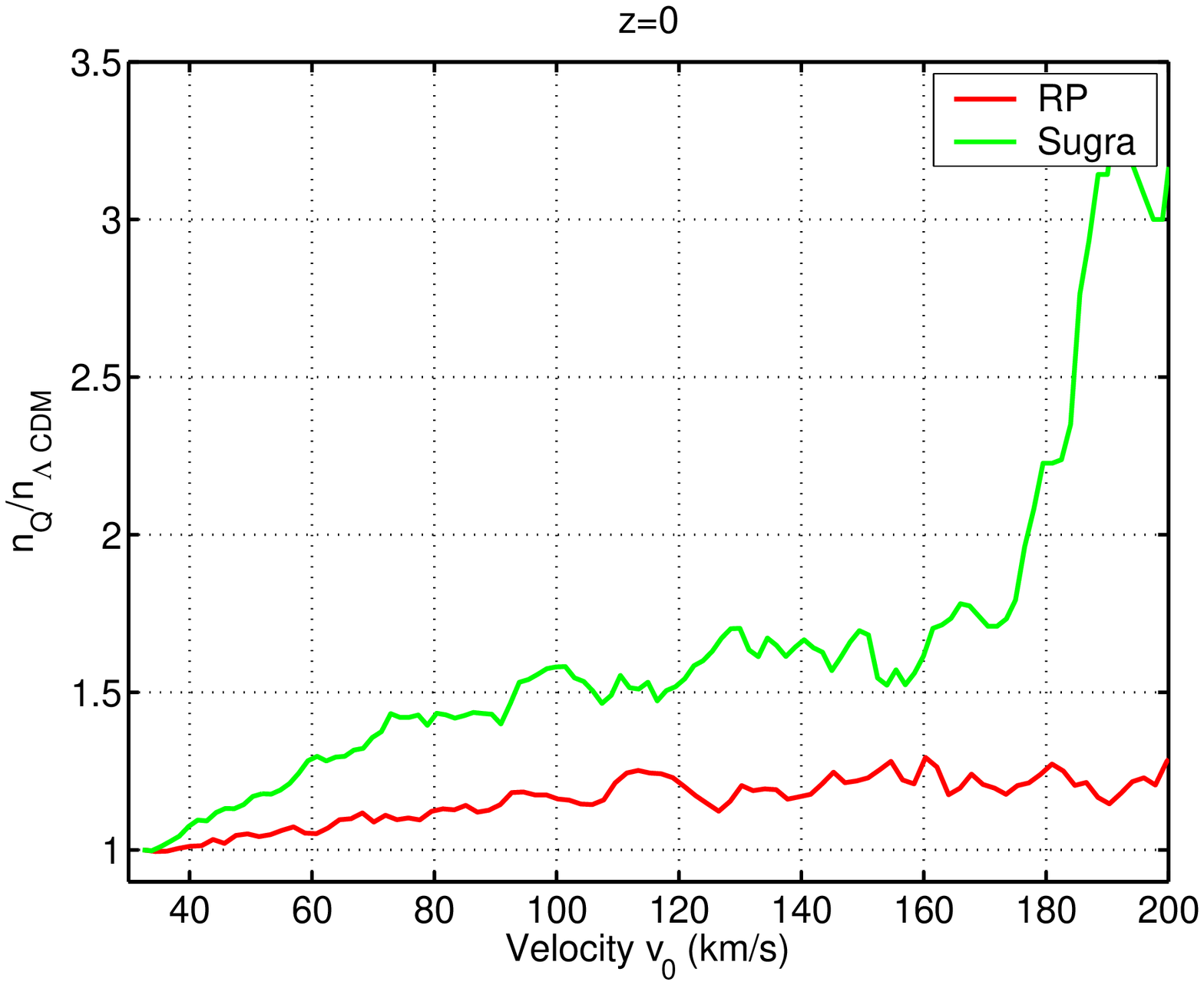} &
\includegraphics[scale=0.3]{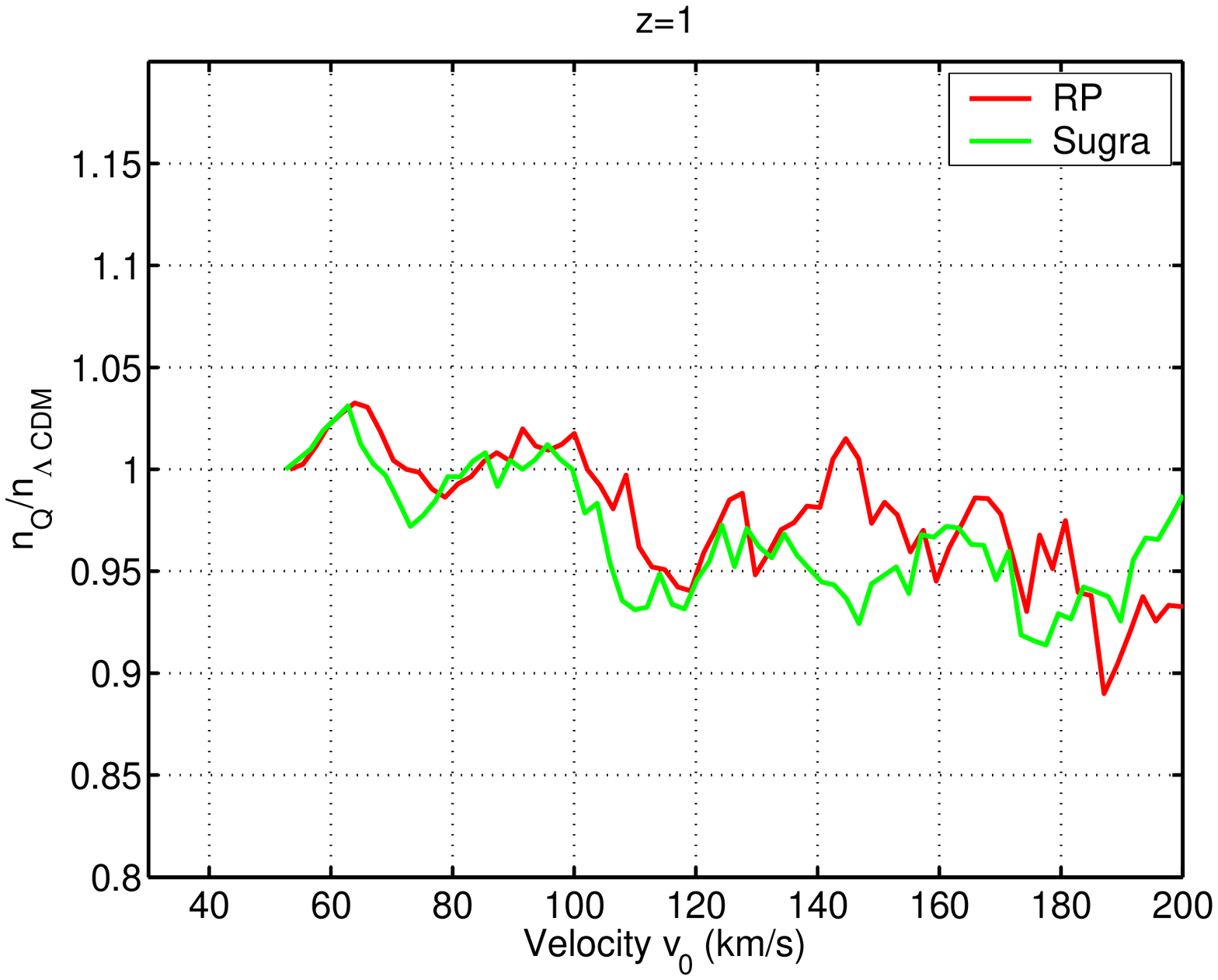}
\end{tabular}
\caption{Number density of halos with internal velocity dispersion larger than a given value $v_0$ in the $\Lambda CDM$ model at $z=0$ (straight line) and $z=1$ (dash-dotted line) (left).
The same number densities for the quintessence models at $z=0$ (center) and $z=1$ (right) are illustrated  in units of the number density of the
$\Lambda CDM$ model.}
\label{disp}
\end{figure}

Our aim is to give an overview of some impacts of quintessence on large-scale structure formation, mainly
on cluster properties like mass functions and internal velocity dispersion. We will treat here realistic
quintessence models that account for the Hubble diagram of type Ia supernovae while other works 
like \cite{klypin,dolag}
made use of toy models for quintessence\footnote{They do not account for supernovae data.} and focused mainly on the cluster mass functions.
We will study dark matter collapse on large-scales through
N-body simulations (Particle-Mesh algorithm).
We will consider $256^3$ particles moving into a box of $32h^{-1}Mpc$ comoving length with $256^3$ grid cells ($h=0.7$),
with the baryon density parameter $\Omega_b$ equal to $4\%$.
At scales lower than $100h^{-1}Mpc$, it has been shown in \cite{brax2} that a $\Lambda CDM$
power spectrum can be used for the initial conditions (the shape of the spectrum in a $\Lambda CDM$ and
in RP and Sugra quintessence models is very similar at those scales). We use the WMAP3 normalisation \cite{wmap3} 
for each of the simulation, with
$rms$ 
density fluctuation level within spheres of $8h^{-1}Mpc$ radius $\sigma_8=0.74$.
The initial redshifts (given in table \ref{chis})
of the simulations are determined such that the linear evolution of the filtered dispersion $\sigma_8$ reaches today this value. 
The linear evolution
of the matter density contrast is given by the solution of the following equation (see \cite{peebles}):
\begin{equation}
\label{dlin}
\dot{a}^2\delta''+\left(\ddot{a}+2\frac{\dot{a}^2}{a}\right)\delta'-\frac{\delta}{a^3}=0,
\end{equation}
where we assumed $a_0=1$, $4\pi m_{Pl}^{-2}\rho_{m,0}=1$ and where a prime and a dot denote a derivative w.r.t. the scale factor $a$ and
time $t$, respectively.
The analytical solution under integral form,
originally derived by Heath in 1977
for universes with matter and cosmological constant, is no longer valid in the case of quintessence, even
with a simple constant eos. We have therefore to solve numerically equation (\ref{dlin}) with initial conditions
given by requiring that the density contrast evolves linearly with the scale factor $a$ after recombination $z\approx 10^2$ 
(matter-dominated universe).
\\
\\
The left plot of Figure \ref{slices} represents a slice of the dark matter 3D density field $\rho_{DM}$ for the best fit $\Lambda CDM$ model at $z=0$.
The other plots in Figure \ref{slices} (center and right) 
illustrate the absolute differences $\rho_{DM}^{S(RP)}/\rho_{DM}^{\Lambda CDM}-1$ 
for the same slice
with respect to the $\Lambda CDM$ model 
for the Sugra and RP models respectively. The bright (dark) regions in those plots 
indicate where the quintessence models are more (less) structured than the $\Lambda CDM$.
We can see that the $\Lambda CDM$ is more clustered in the filaments than
the quintessence models (dark regions in the central and right parts of Figure
\ref{slices}) leaving the last more structured outside the filaments (bright regions). 
Let us now examine the discrepancies in the effect of quintessence or cosmological constant 
on dark matter halos. Here, they are located 
thanks to the Friend-Of-Friend algorithm, which groups inside a same cluster all the particles 
that are within a certain fraction of the mean distance between particles from each other. 
We will consider clusters of at least $100$ particles
and a reference distance used to identify the cluster of $20\%$ of the mean distance between particles.
Figure \ref{massfunc} (left) represents the number density 
of clusters with a mass above a given value $M_0$ at $z=0$ and $z=1$ for the $\Lambda CDM$ models, while
the central and right part of Figure \ref{massfunc} illustrate the ratio between the number density in models with quintessence
and the number density of $\Lambda CDM$. The differences are of the order of $50\%$ for high masses, even at $z=0$. In the toy models
used in \cite{klypin}, the mass functions at $z=0$ were found almost undistinguishable. Therefore, 
using the right cosmological parameters
for quintessence models, as suggested by distance-redshifts measurements, increases the differences on the mass functions at $z=0$
found in \cite{klypin}
for toy models. Another important difference with \cite{klypin} is the following:
the authors also found that the RP model was more clustered than the Sugra which was more clustered than the $\Lambda CDM$. 
This order is, however,
due to their assumption of same cosmological parameters which yields that the differences in structure formation
are only due to different cosmic expansion.
This order reflects the acceleration provided by quintessence : their RP model accelerates less ($\omega_{RP}\approx -0.4$) than their Sugra model
($\omega_{Sugra}\approx -0.8$) and the cosmological constant $\Lambda CDM$ ($\omega_{\Lambda}=-1$). Here, we find that the $\Lambda CDM$
model is more clustered than the RP and Sugra quintessence models, and this fact is due to different cosmic expansion and the 
different cosmological parameters required by distance-redshifts measurements.
\\
\\
Let us now examine the internal velocity dispersion of dark matter halos, given by $\sigma_{int}^2=1/N\sum_{i=1}^{N}||\overrightarrow{v}_i-\overrightarrow{v}_{CM} ||^2$
(with $N$ the number of particles in a given cluster and $\overrightarrow{v}_{CM}$ the center-of-mass velocity).
This is done in Figure \ref{disp}: on the left the number density of of halos with velocity dispersion higher than
a given value is given for the $\Lambda CDM$ model at $z=0$ and $z=1$. On the center and right part of the figure,
the number densities for quintessence models are given in units of the $\Lambda CDM$ amount. Quintessence 
lead to higher internal velocity dispersion than the $\Lambda CDM$ at $z=0$. This can be 
justified by the fact that these models produce more
lighter halos at $z=0$ which are less virialised. At $z=1$, the differences with a $\Lambda CDM$ are smaller, which
means that more light halos are formed in quintessence models between $z=0$ and $z=1$ (see also Figure \ref{massfunc}).
\section{Conclusions}
In this paper, we have illustrated some impacts of quintessence on the distribution of dark matter on large-scales.
We have shown through to an analysis of Hubble diagram of type Ia supernovae
that the cosmological parameters $(\Omega_{m,0}, \Omega_{Q,0})$ for quintessence models are quite different
than in the concordance $\Lambda CDM$ model, as might be expected from the different cosmic acceleration
quintessence models provide. 
Cosmological models with quintessence are spatially closed
and exhibit a slightly smaller amount of matter than the concordance model with cosmological constant.
The effect of the necessary different cosmological parameters was underestimated in the litterature 
on the imprints of quintessence in structure formation.
We have shown that this lead to more important deviations between the predictions of a $\Lambda CDM$ and quintessence models at $z=0$.
As well, considering the same cosmological parameters for quintessence and cosmological constant does not fit Hubble diagram
data but it also introduces a bias in the interpretation of the imprints of quintessence on the large-scale distribution of dark matter. With the same
cosmological parameters, quintessence models were found in \cite{klypin} to produce more structures and to be more clustered than a $\Lambda CDM$ 
scenario. However, once the correct cosmological parameters are taken into account, we have shown that this is exactly the opposite.
In this paper, we have shown that quintessence yields to more structures outside the filaments, lighter halos with higher internal
velocity dispersion. The present study illustrates how dark matter is sensitive to an expansion driven by
quintessence or cosmological constant.
These differences on the dark matter collapse, even at $z=0$, will lead to observable imprints on
galaxy formation. An interesting point for further works will be to establish 
constraints on the nature of dark energy from galaxy properties through numerical simulations.

\label{lastpage}


\begin{thebibliography}{99}
\bibitem{riess1} Riess A.G. et al., ApJ 116 (1998), 1009.
\bibitem{perlmutter} Perlmutter S. et al., ApJ 517 (1999), 565.
\bibitem{riess2} Riess A.G. et al., ApJ 607 (2004), 665-687.
\bibitem{snls} P. Astier et al., Astronomy and Astrophysics, 2005.
\bibitem{bennett} Bennett C.L. et al., ApJ Suppl. 148 (2003) 1.
\bibitem{tegmark1} Tegmark M., ApJ 514 (1999), L69.
\bibitem{tegmark2} Tegmark M. et al., Phys.Rev. D69 (2004) 103501.
\bibitem{allen} Allen S.W., Schmidt R.W. \& Fabian A.C., MNRAS 334 (2002) L11, astro-ph/0205007.
\bibitem{mohayaee} Mohayaee R., Tully R. B., Astrophys.J. 635 (2005) L113-L116.
\bibitem{weinberg} Weinberg S., Rev. Mod. Phys. 61, 1 (1989)
\bibitem{padmanabhan} Padmanabhan T., Class. Quantum Grav. 22 17 (2005) L107-L112
\bibitem{ratra}  Ratra B. \& Peebles P.J.E., Phys. Rev. D37 (1988) 3406.
\bibitem{brax}  Brax P. \& Martin J., Phys.Rev. D61 (2000) 103502, astro-ph/9912046.
\bibitem{barreiro}  Barreiro T., Copeland E.J. \& Nunes N.J., Phys. Rev. D61 (2000) 127301, astro-ph/9910214.
\bibitem{frieman} Frieman J.A. et al., Phys.Rev.Lett. 75 (1995)
2077.\\  Frieman J.A. \& Waga I., Phys.Rev. D57 (1998) 4642.
\bibitem{steinhardt}   Steinhardt P.J., Wang L. \& Zlatev I., Phys.Rev. D59 (1999) 123504,
astro-ph/9812313. \\
Zlatev I., Wang L. \& Steinhardt P.J., Phys.Rev.Lett. 82 (1999)
896-899, astro-ph/9807002.
\bibitem{dipietro} Di Pietro E. \& Claeskens J.-F., M.N.R.A.S.  341 (2003) 1299, astro-ph/0207332.
\bibitem{caresia} Caresia P., Matarrese S. \& Moscardini L., ApJ 605 (2004) 21-29.
\bibitem{brax2} Brax P., Martin J. \& Riazuelo A., Phys.Rev. D62 (2000) 103505, astro-ph/0005428.
\bibitem{ma} Ma C.P. et al., ApJ 521 (1999) L1-L4.\\
 Ma C.P., ApJ 508 (1998) L5-L8.
\bibitem{bode} Bode P. et al., ApJ 551 (2001) 15-22.
\bibitem{lokas} Lokas E.L., Bode P. \& Hoffman Y., Mon.Not.Roy.Astron.Soc. 349 (2004).
\bibitem{munshi} Munshi D., Porciani C. \& Wang Y., Mon.Not.Roy.Astron.Soc., 349, 281 (2004)
\bibitem{benabed} Benabed K. \& Bernardeau F., Phys.Rev. D64 (2001) 083501
\bibitem{dolag} Dolag K. et al., Astron. Astrophys. 416 (2004) 853-864, astro-ph/0309771.
\bibitem{klypin} Klypin A. et al., ApJ 599 (2003) 31-37. \\
Mainini R., Maccio A.V., Bonometto S.A., Klypin A., Astrophys.J. 599 (2003) 24-30
\bibitem{solevi} P. Solevi et al., Mon. Not. Roy. Astron. Soc. 366 (2006) 1346-1356.
\bibitem{peebles} Peebles P.J.E., Principles of Physical Cosmology, Princeton University Press, 1993.
\bibitem{wmap3} D.N. Spergel et al., astro-ph/0603449
\end{thebibliography}
\end{document}